\documentclass[%
 reprint,
 amsmath,amssymb,
 aps,
]{revtex4-2}
\usepackage{amsmath}
\usepackage[toc]{appendix}
\usepackage{graphicx}
\usepackage{dcolumn}
\usepackage{comment}
\usepackage{bm}
\usepackage{float}
\usepackage{xcolor}

\usepackage{booktabs}

\newcommand\scalemath[2]{\scalebox{#1}{\mbox{\ensuremath{\displaystyle #2}}}}

\begin{document}
\preprint{APS/123-QED}

\title{True muonium resonant production at $e^+e^-$ colliders with standard crossing angle}

\author{Ruben Gargiulo}%
 \email{ruben.gargiulo@lnf.infn.it}
\affiliation{Università degli Studi La Sapienza, Piazzale Aldo Moro 5, 00185 Roma, Italy}%

\author{Elisa Di Meco}
 \affiliation{INFN Laboratori Nazionali di Frascati, Via Enrico Fermi 54, 00044 Frascati, Italy}
\author{Daniele Paesani}%
\affiliation{Università degli Studi di Roma Tor Vergata, Via della Ricerca Scientifica 1, 00133 Roma, Italy}%
\author{Stefano Palmisano}%
\affiliation{Università degli Studi La Sapienza, Piazzale Aldo Moro 5, 00185 Roma, Italy}%
\author{Eleonora Diociaiuti}
 \affiliation{INFN Laboratori Nazionali di Frascati, Via Enrico Fermi 54, 00044 Frascati, Italy}
\author{Ivano Sarra}
 \affiliation{INFN Laboratori Nazionali di Frascati, Via Enrico Fermi 54, 00044 Frascati, Italy}
\date{\today}

\begin{abstract}
True muonium ($\mu^+\mu^-$) is the heaviest and smallest bound state not containing hadrons, after true tauonium ($\tau^+\tau^-$) and mu-tauonium ($\mu^\pm\tau^\mp$). 
One of the proposed methods to observe the spin 1 fundamental state of TM, which has the smallest lifetime among TM spin 1 states, was to build an $e^+e^-$ collider with a large crossing angle ($\theta \sim 30^\circ$) in order to provide TM with a large boost and detect its decay vertex in $e^+ e^-$.
The following paper will instead show that TM excited states can be observed in relatively large quantities ($\mathcal{O}$(10)/month) at a $e^+e^-$ collider with standard crossing angle, after setting their center-of-mass energy to the TM mass ($\sim2m_{\mu}=211.4$ MeV).

\end{abstract}

\maketitle


\section{\label{sec:intro} Introduction}
Quantum electrodynamics (QED) predicts the existence of several bound states, in addition to standard atoms, such as purely leptonic systems.
Given the absence of clear signals beyond the Standard Model (BSM), precision measurements of QED bound states might be employed as new physics probes.
However, observable quantities of bound states containing hadrons have large theoretical uncertainties from unknown non-perturbative quantum chromodynamics effects, while the properties of purely leptonic bound states (such as positronium \cite{KARSHENBOIM_2004}) can be calculated very precisely. 
At the same time, purely leptonic bound states containing electrons are limited in their BSM discovery potential through atomic spectroscopy by the mass suppression due to the small term $m_e/\Lambda_{BSM}$. In contrast, bound states containing only $\mu$ and $\tau$ particles have much larger reduced masses so their BSM sensibility is enhanced \cite{refId0}. 
One of the possible bound-state choices is represented by the so-called True Muonium (TM), a bound state containing a $\mu^+$ and a $\mu^-$, that, with its 211.4 MeV mass and 512 fm Bohr radius constitutes the heaviest and smallest purely leptonic QED atom right after true tauonium ($\tau^+\tau^-$) \cite{TT} and mu-tauonium ($\mu^\pm\tau^\mp$). \\
It should be noted that addressing the search for BSM signals to muons is reasonable because of the long-standing ``muon problem": the coincidence that multiple observables in the muonic sector deviate from either theoretical predictions or similar results with other leptonic flavours \cite{mp1}.
TM precision measurements are also useful to the Standard Model itself, in the hypothesis of the absence of new physics at accessible scales, because its hyperfine splitting shifts are directly sensible to contributions from hadronic vacuum polarization \cite{PhysRevA.95.012505}, like for the muon anomalous magnetic moment \cite{crivellin}.\\
It is interesting to point out that the $(\mu^+\mu^-)$ bound state is called ``true muonium" since the name ``muonium" was previously used for the $\mu^+e^-$ bound state. Indeed, the first studies of TM only began as its production was shown to be feasible. Positronium and muonium have been observed and studied, while TM has not been observed yet. \\
Several production mechanisms have been proposed for TM, including meson decays, like $\eta \to TM\gamma$ \cite{PhysRevD.99.033008} \cite{lhcb} and $K_L \to TM\gamma$ \cite{PhysRevD.98.053008}, or electron-nucleus $eZ \to e \,TM\,Z$ \cite{Holvik:1986ty}, nucleus-nucleus $Z_1 Z_2 \to Z_1 Z_2 \,TM $ \cite{Ginzburg:1998df} and electron-positron $e^+e^- \to TM (\gamma)$ \cite{Brodsky:2009gx} collisions. 
This work focuses on the last method. 

\section{True Muonium properties}
\begin{figure}[H]
\centering
    \includegraphics[width=\columnwidth]{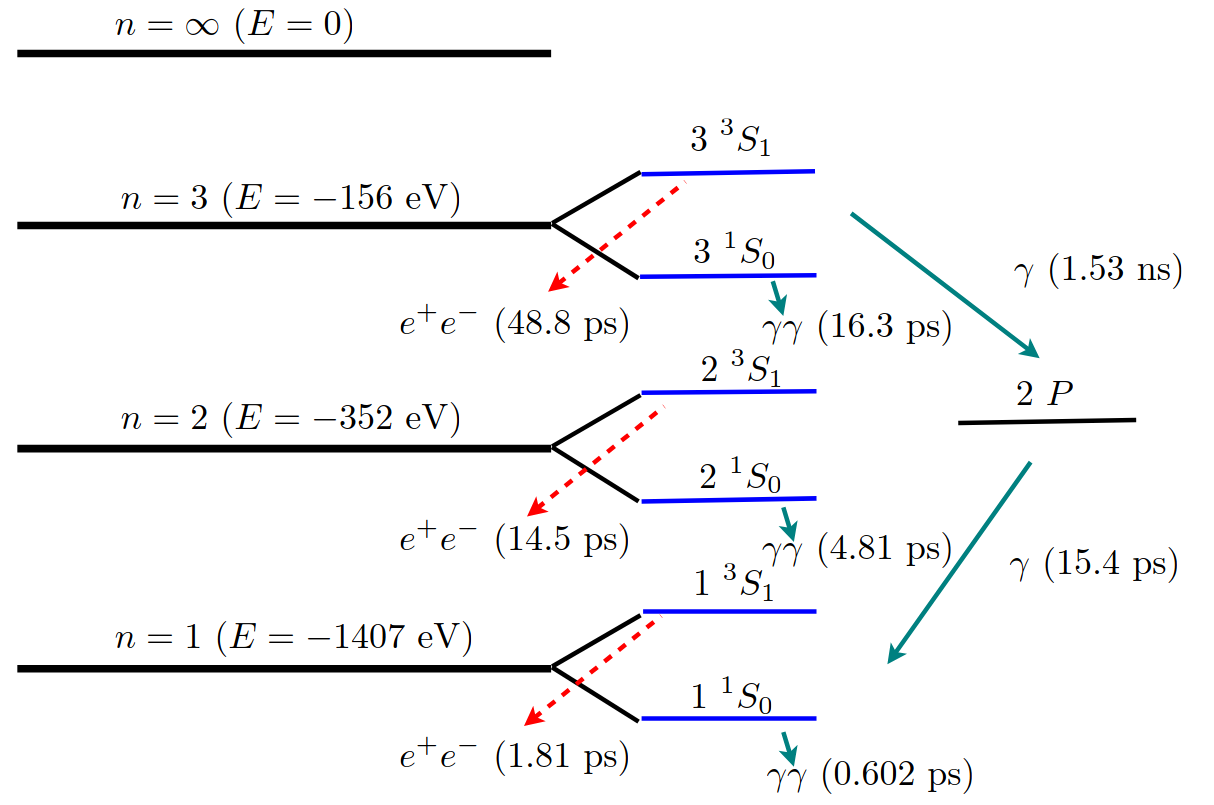}
    \caption{True muonium levels, lifetimes and transitions diagram for $n\leq 3$ (spacing not to scale)~\cite{Brodsky:2009gx}.}
    \label{fig:TM}
\end{figure}

The TM energy levels can be calculated by rescaling the positronium spectrum: the binding energy of the deepest level (1S) was evaluated to be $\mathrm{B.E.(1S)} = 1.4$ keV, as shown in Figure \ref{fig:TM}.\\
Like positronium, TM has two spin states: para-TM (spin 0), which decays to $\gamma \gamma$, and orto-TM (spin 1), which decays to $e^+e^-$.\\
For each of the two spin states, spontaneous transitions from the $(n+1)S/P$ to the $nP/S$ are possible. 
While for S states these transitions compete with the decay, for P states they are mandatory. Indeed, the ortho-$P$ states have $P=\left(-1\right)^{l+1=2}=1$ and $C=\left(-1\right)^{l+s=1+1=2}=1$, so they cannot annihilate to $e^+e^-$ via a photon in the s-channel as the $S$ states because of parity and charge conjugation conservation. The lifetimes of the n-th $S$ levels for the two spin states $s=0, \, 1$ are proportional to $n^3$ (at lowest order), as follows:
\begin{align}
    \tau(nS_{s=1} \to e^+e^-) = \frac{6 \hbar n^3}{\alpha^5 m_\mu c^2} \sim n^3 \times 1.8 \, \mathrm{ps} \\
    \tau(nS_{s=0} \to \gamma \gamma) = \frac{1}{3} \tau(nS_{s=1} \to e^+e^-)\,.
\end{align}
These lifetimes are much smaller than the muon lifetime, therefore the muons inside TM can be assumed as stable particles.

\section{True muonium production in $e^+e^-$ collisions}
Once known the main TM characteristics, it is possible now to focus on its production methods. 
The most abundant processes to produce TM from  $e^+e^-$ collisions are: off-resonance $e^+e^- \to \mathrm{TM }\gamma$ interactions at $\sqrt{s} > 2m_\mu$  with a hard recoil photon, and resonant $e^+e^- \to \mathrm{TM}$ interactions at $\sqrt{s} \sim 2m_\mu$ \cite{Brodsky:2009gx}. 
Off-resonance interactions have a cross-section of:
\begin{equation}
   \sigma_{OFF\,R.} \sim \frac{\pi}{2} \left [ \mathrm{ln} \left( \frac{1 + c_0}{1- c_0} \right) -c_0  \right] \frac{\alpha^6}{s}\,,
\end{equation}
where $c_0$ is the cosine of the detector acceptance polar angle,
while resonant interactions have a much 
larger cross-section of:
\begin{equation}\label{res}
    \sigma_{ON\,R.} = 2\pi^2 \frac{\alpha^3}{s} = \frac{ \pi^2 \alpha^3}{2 m_{\mu}^2} = 66.6 \, \mathrm{nb}\,.
\end{equation}


It must be pointed out that the probability to produce TM in a state $n$ is proportional to $n^{-3}$ \cite{Brodsky:2009gx}, and the normalization factor is $\zeta(3)$, where:
\begin{equation}
    \zeta(k) = \sum_{n=1}^{+\infty} \frac{1}{n^k}
\end{equation}
is the Riemann Zeta function.

For TM production at colliders, the cross-section in eq. \eqref{res} is reduced accounting for the the probability $p$ that the beam center-of-mass energy is in the energy window $(m_\mu - B.E.(1S), \, m_\mu)$ where bound states are allowed \cite{Brodsky:2009gx}.


Considering that the energy window width 
$\Delta E =  B.E.(1S) = 1.4$ keV is much smaller than the beam's energy spread $\sigma_E$, for a Gaussian distribution this $p$ factor is simply given by the peak value of the $\sqrt{s}$ probability density function (where $\sigma_{\sqrt{s}} = \sqrt{2} \sigma_E$) multiplied by $\Delta E$, namely

\begin{equation}
    p = \frac{\Delta E}{2 \sqrt{\pi} \sigma_E} =  \frac{\Delta E}{2\sqrt{\pi}m_\mu} \left(  \frac{\sigma_E}{E} \right)^{-1} = 3.7 \times 10^{-6} \left(  \frac{\sigma_E}{E} \right)^{-1}\,.
    \label{pfact}
\end{equation}

\section{True muonium production at DA$\Phi$NE}
\label{dafnetm}
Existing proposals for TM observation using resonant interactions involve the construction of an  $e^+e^-$ collider with a very large crossing angle $\theta \sim 30^\circ$ so that TM has enough boost to allow the observation of its 1S decay vertex without requiring impractically small uncertainties on vertex and interaction point positions \cite{dimus} \cite{russi}.\\ 
In the following it will be shown instead that, by relaxing the requirement to observe the fundamental state and limiting to the study of excited states, TM can be discovered at 5$\sigma$ at existing $e^+e^-$ colliders, such as DA$\Phi$NE in Frascati \cite{dafne}, if running at the proper center-of-mass energy. It is indeed possible to exploit its non-zero crossing angle and provide enough boost to TM excited states, therefore observing their decay vertices and discriminating the signal over the background. It will be shown that, by running DA$\Phi$NE at a center-of-mass energy equal to $\sqrt{s} = 2m_{\mu} = 211.4$ MeV, the discovery of TM excited states with significance exceeding 5$\sigma$ is possible with one month of data taking, using a cylindrical detector interface that embeds a multi-layer silicon tracker, a high-granularity electromagnetic calorimeter and a cosmic ray veto.\\
The technical difficulties of operating DA$\Phi$NE or other colliders with 105.7 MeV beams are beyond the scope of this article. As a proof-of-concept, a hypothetical discovery experiment using DA$\Phi$NE machine parameters will be described.
The use of the DA$\Phi$NE collider as a benchmark for TM production is reasonable considering that, currently in the world, there is no other \rm{$e^+e^-$} machines operating at such low energies.
It should also be highlighted that the performances of the DA$\Phi$NE collider employed in the following are the ones delivered to the KLOE-II experiment at the nominal center-of-mass energy of 1020 MeV. Moreover, the details of the beam pipe and the interaction chamber will be also taken from the ones of the KLOE-II experiment at DA$\Phi$NE \cite{kloe}.

With a relative beam energy spread of $2 \times 10^{-3}$ \cite{bes}, the $p$-factor from eq. \eqref{pfact} is $ 1.85 \times 10^{-3}$, leading to a realistic production cross-section of:
$$\sigma_{ON\,R.}^{real} \sim 124\,\text{pb}\,$$ 
therefore, with a daily luminosity of 10 pb$^{-1}$/day \cite{catia}, $\sim$1240 TM atoms are produced per day. 

 Note also that with here described resonant interactions, the TM is created in the spin 1 state, because its production is mediated by a virtual hard photon in the s channel, so it decays mostly
to $e^+e^-$ \cite{Brodsky:2009gx}.\\
The DA$\Phi$NE crossing angle between the electron and positron beams is $\theta=50$ mrad, thus producing a boost in the radial $x$  direction \cite{catia}. 
The TM boost in $x$ is then $m_\mu \sin{\theta} = 5.3$ MeV, hence $\beta_x \gamma = 2.5 \times 10^{-2}$, resulting in the average path lengths of TM states shown in Table \ref{tab:lifetimes}.
The fluctuation of the boost in the Z direction is $\sqrt{2} m_{\mu} \sigma_E / E \sim 300 $ keV, about 18 times less than the nominal boost in $x$, so it is safe to assume that the TM only moves in the $x$ direction.
\begin{table}[H]
\centering
\begin{tabular}{@{}c|ccc@{}}
\toprule
\multicolumn{1}{l|}{\textbf{\textbf{$n$}}} &
  \multicolumn{1}{c}{\textbf{\textbf{Relative yield}}} &
  \multicolumn{1}{c}{\textbf{\quad\textbf{$\tau_n$ [ps]}\quad}} &
  \textbf{\quad\textbf{$l_n$ [$\mathrm{\mu}$m]}\quad} \\ \midrule
1  & 0.83333 & 1.8    & 13.6    \\
2  & 0.10417 & 14.5   & 108.4   \\
3  & 0.03086 & 48.8   & 366.0   \\
4  & 0.01302 & 115.7  & 867.6   \\
5  & 0.00667 & 225.9  & 1694.4  \\
6  & 0.00386 & 390.4  & 2928.0  \\
7  & 0.00243 & 619.9  & 4649.6  \\
8  & 0.00163 & 925.4  & 6940.4  \\
9  & 0.00114 & 1317.6 & 9882.0  \\
10 & 0.00083 & 1807.4 & 13555.6 \\ \bottomrule
\end{tabular}%
\caption{Relative yield, lifetime and average path length (assuming $\beta \gamma = 2.5 \times 10^{-2}$) for TM spin 1 states, as a function of $n$.}
\label{tab:lifetimes}
\end{table}
The path length values in the table should be compared to the interaction point $x$ position uncertainty $\sigma_X$, where $(\sigma_X, \sigma_Y, \sigma_Z)=(200 \, \mathrm{\mu m}, 2.6 \, \mathrm{\mu m}, 20 \, \mathrm{mm})$ \cite{catia}.
\subsection{Initial State Radiation}
\label{sectisr}
The effects of ISR (initial state radiation) on True Muonium production should be carefully evaluated. On the one hand, the center of mass energy for the hard, partonic collision is reduced by radiation, which leads to a reduction in the cross section for the production of TM. On the other hand it boosts the electron pair, resulting in a boost of the intermediate TM.

To address the first effect the partonic cross section for producing TM is assumed to be constant and equal to Eq.\eqref{res} within the window $\left[2 m_\mu - \Delta E, 2 m_\mu \right]$, while it goes to zero outside this range. Then if $\mathcal{G}_{\text{BES}}(s)$ is the Gaussian function describing the beam energy spread and $f_{\text{ISR}}(x; s)$ is the radiator function, expressing the probability of an initial electron pair to carry a fraction $x$ of the center of mass energy (see Appendix \ref{app:isr}), the effective cross-section reads
\begin{equation}\label{eq:isr-rate}
\scalemath{0.95}{\sigma_{\text{TM,eff.}} = \int d\, s' \,\mathcal{G}_{\text{BES}}(s') \int
d\, x \, f_{\text{ISR}}(x; s') \sigma_{\text{TM}}(x^2 s')}\,,
\end{equation}
where the $x$ integral is evaluated with the following extrema:
\begin{eqnarray}
x_{\text{min}}(s') = \min\left[1, \frac{2 m_\mu - \Delta E}{\sqrt{s'}}\right]
\\
x_{\text{max}}(s') = \min\left[1, \frac{2 m_\mu}{\sqrt{s'}}\right]\,.
\end{eqnarray}
By evaluating the integral numerically, a $\sigma_{\text{TM,eff.}}$ of $93.9$ pb is obtained.

\begin{figure}[H].
\centering
    \includegraphics[width=\columnwidth]{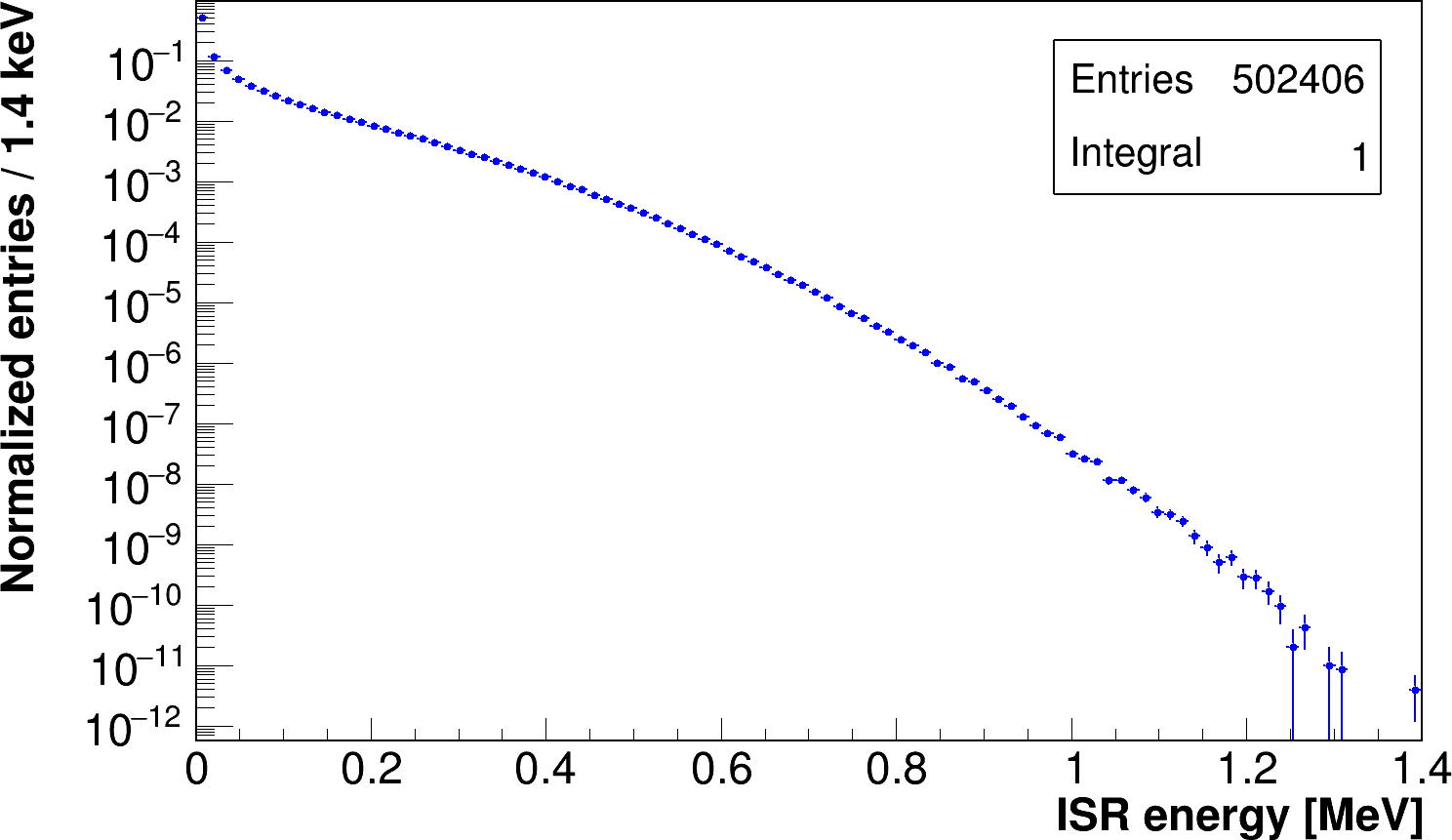}
    \caption{Distribution of energy radiated away by ISR for TM production.}
    \label{fig:isr}
\end{figure}

Concerning the TM, random values of the energy of each collision were extracted according to the beam energy spread distribution and then the probability that, after ISR, the partonic center of mass energy is within the TM production window, was computed as
\begin{equation}\label{eq:isr-boost-weight}
\int_{x_{\text{min}}(s')}^{x_{\text{max}}(s')} d\,x f_{\text{ISR}}(x; s').
\end{equation}
A distribution of the energy carried away by ISR, which can be approximated to the energy of a single ISR photon, is then obtained and shown in Figure \ref{fig:isr}.

\section{Backgrounds discussion}
At a center-of-mass energy of $\sqrt{s} = 2m_\mu \sim 211.4$ MeV, the only particles that can be produced are electrons, photons, single neutral pions, and muon pairs. Charged pions are excluded because there is not enough invariant mass available to produce a pair. A single charged pion must indeed be accompanied by a single electron to conserve charge, thus violating the lepton number conservation. Note that none of the remaining particles have lifetimes comparable with the one of TM, because electrons and photons are stable, neutral pions decay in $8.5  \times 10^{-17}$s, and muons, that never decay to $e^+e^-$, have a lifetime of $2.2\,\mathrm{\mu}$ s.  Therefore, any displaced decay vertex in $e^+e^-$ is a sign of TM presence, meaning that the only background is given by fake $e^+e^-$ displaced vertices, due to particle mis-identification or to the finite resolution on the  reconstructed vertex position. 
A detailed discussion of the backgrounds will follow.

\subsection{Bhabha scattering}
\begin{figure}[H].
\centering
    \includegraphics[width=\columnwidth]{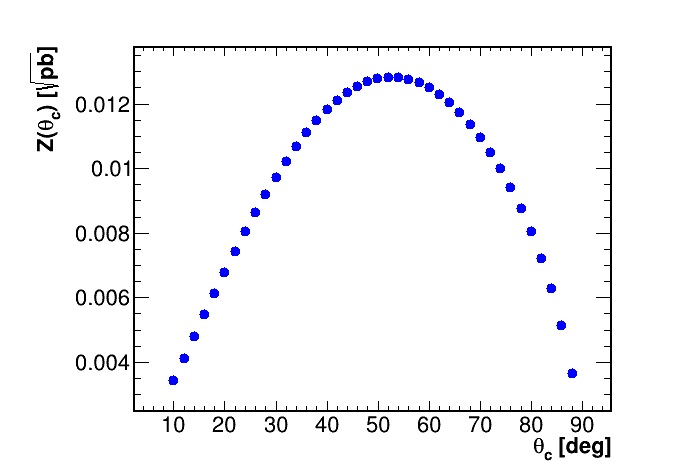}
    \caption{Significance scan in detector acceptance angle $\theta_c$, at $1^\circ$ steps (see eq. \eqref{signtheta}). The peak is around $53^\circ$.}
    \label{fig:theta}
\end{figure}

The dominant background is represented by Bhabha scattering, whose yield is several orders of magnitudes above the signal.
Bhabha scattering produces electron pairs with the same energy as TM decay products, and it must be suppressed with appropriate cuts based on decay vertex and tracks reconstruction, in particular the tracks polar angle.\\
The differential Bhabha scattering cross section, at lowest order, is:
\begin{small} 
\begin{equation}
\label{bhabhaeq}
    \dfrac{d\sigma}{d\Omega} = \dfrac{\alpha^2}{2s} \left[\dfrac{1 + \cos^4(\theta/2)}{\sin^4(\theta/2)} - \dfrac{2\cos^4(\theta/2)}{\sin^2(\theta/2)} + \dfrac{1 + \cos^2(\theta)}{2}\right]\end{equation}
    \end{small} 
Electron pairs originating from Bhabha scattering thus have predominantly small $\theta$ angles, while
TM decay products are distributed as
$\dfrac{dN}{d\cos{\theta}} \propto \left( 1 + \cos^2{\theta} \right)$ (neglecting the effect of the small TM boost).
Therefore, an angular cut [$\theta_c$,\,$\pi - \theta_c$] can be efficiently used to partly discriminate signal over background, using the asymptotic significance \cite{pratstat}:
\begin{equation}
    Z(\theta_c) = \frac{\sigma_{\mathrm{TM}}(\theta_c < \theta < \pi - \theta_c)}{\sqrt{\sigma_{\mathrm{Bhabha}}(\theta_c < \theta < \pi - \theta_c)}},
    \label{signtheta}
\end{equation}
shown in Figure \ref{fig:theta}, as a figure of merit. The shape of $Z(\theta_c)$ does not change if the signal or background yields are modified by other cuts independent from $\theta_c$, so its maximum can be used to establish the optimal angular cut. As a trade-off between such optimization ($\theta^c_{\text{opt}} = 53^\circ$) and a feasible detector geometry an angular cut of $\theta_c = 60^\circ$ is applied, therefore the TM production cross section reduces to $\sigma_{ON\,R.}^{real}(\theta_c=60^\circ) \sim 39$ pb (taking ISR into account), corresponding to 390 TM/day at 10 pb$^{-1}$/day, while the Bhabha cross section at lowest order is $\sigma_{\mathrm{Bhabha}}(\theta_c=60^\circ) \sim 9.35 \, \mathrm{\mu}$b.
This angular cut is not enough to efficiently suppress the Bhabha scattering background, hence other cuts based on tracks and vertex reconstruction are necessary.\\
The precision on tracks reconstruction could be compromised by the multiple Coulombian scattering in the interaction region, that for this case (as in the KLOE-II experiment) is composed by a 50 $\mu$m thick beryllium beam pipe, surrounded by a 500 $\mu$m thick AlBeMet spherical vacuum chamber vessel \cite{cpt}. 
 In order to completely avoid the multiple Coulombian scattering in the vacuum chamber wall, the multi-layer silicon tracker must be installed around the beam pipe and inside the vacuum chamber.

In the following, it will be shown how a mitigation strategy is able to reject the Bhabha scattering events with an analysis based on reconstructed vertex position and the so-called Line Of Response (LOR), derived from the Positron Emission Tomography \cite{pet}.

\begin{figure}[H]
    \includegraphics[width=0.9\columnwidth]{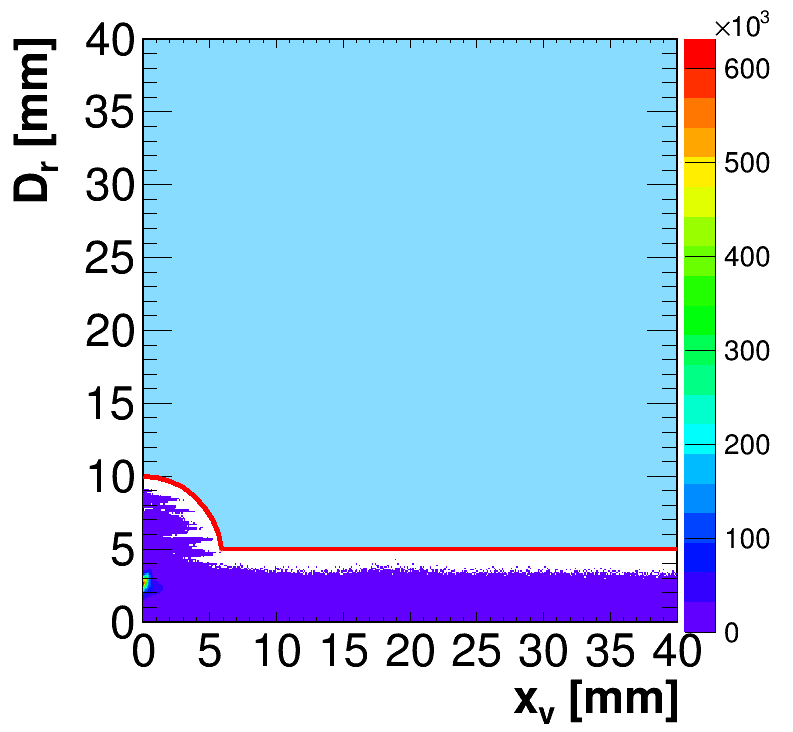}
    \caption{Joint distribution of $D_r$ and $x_v$ for $5 \times 10^{9}$ Bhabha scattering events, after the cut on $\alpha$. The cuts $D_r > 5$ mm and the circular one are represented in red, and the signal region is filled in cyan.}
    \label{fig:bhabhath2}
\end{figure}

A proof-of-concept simulation was performed to understand the response to Bhabha scattering events at $\sqrt{s} = 2m_\mu$ of a cylindrical multi-layer silicon tracker with 100 $\mu$m thick layers (here only 2 for simplicity) and 10 $\mu$m spatial resolution. The simulation geometry embeds:
\begin{itemize}
    \item the cylindrical 50 $\mu$m thick beryllium beam pipe placed at a 4.4 cm radius,
    \item a cylindrical 100 $\mu$m thick silicon layer placed at a 5 cm radius,
    \item a second cylindrical silicon layer at a 7 cm radius,
\end{itemize}
and its physical model takes into account:
\begin{itemize}
    \item a $10 \mu$m spatial resolution on both $r\phi$ and $z$,
    \item the multiple Coulombian scattering in all materials,
    \item the boost due to the non-zero crossing angle,
    \item uncertainties on the interaction point,
    \item the beam energy spread.
\end{itemize}
For each event, the vertex position is reconstructed as the midpoint of the two closest approach points of the lines extrapolated from tracks hits in the silicon layer. As explained before, an additional quantity, $D_r$, is calculated, representing the distance between the beamline and the LOR connecting the two hits from the $e^+$ and $e^-$ tracks in the first silicon layer.
Using parallel computing, about $N_{\text{Bhabha}} = 5 \times 10^{9}$ Bhabha scattering events, including the emission of photons, were simulated using BabaYaga$@$NLO \cite{babayaga}, a proper next-to-leading order event generator, and reconstructed as described above.
The Bhabha scattering cross section was re-evaluated via BabaYaga$@$NLO, getting 9.0 $\mu$b, so it was only sligthly modified with respect to the value obtained at the leading order with the same angular cut.
Given that the emission of hard photons makes the angle between the charged tracks smaller, a cut $\alpha > 177^o$ on the opening angle between the electron and the positron has been applied. Indeed, the signal has a minimum opening angle of $\alpha = 2 \sin^{-1}{\frac{m_{TM}}{\sqrt{m_{TM}^2 + p_x^2}   }  } = 177.13^\circ$, where $p_x = 5.3$ MeV is the TM momentum (see sect. \ref{dafnetm}), therefore the signal efficiency of this cut is approximately 1.
The joint distribution of $D_r$ and $x_v$ after the cut on $\alpha$ is shown in Figure \ref{fig:bhabhath2}. By applying a cut $D_r > 5$ mm and a circular cut with the formula:
\begin{equation}
    \sqrt{(D_r - 4\,\text{mm})^2 + x_v^2} > 6 \, \text{mm}
    \label{eq:circ}
\end{equation}
no event is left in the signal region from the distribution in Figure \ref{fig:bhabhath2}.
The probability that a Bhabha scattering event enters the signal region is then less than $1/N_{\text{Bhabha}} = 2 \times 10^{-10}$.
With the three cuts sketched above ($\alpha > 177^o$, $D_r > 5$ mm, and the circular one) the background from Bhabha scattering events is reduced to 0.54 events per month.

Note that, in a fraction of radiative Bhabha scattering events, the $x_v$ and $D_r$ based background rejection can be worsened, if the emitted photons interact before or inside the silicon tracker.
Photons not collinear to one of the tracks are generally vetoed by the calorimeter, while collinear photons interacting before it can be rejected by using a multi-layer silicon tracker embedding highly granular pixel sensors, featuring a spatial resolution of $\sim$10$\mu$m. The tracker can indeed discriminate, by geometrically distinguishing the associated hits, the passage of electrons produced by Compton scattering, or $e^+e^-$ pairs created by photon conversions, in addition to the two original back-to-back charged tracks.

A proper detector design, optimizing for instance the number of layers and the distances between them, would suppress this type of events at small enough levels to not sensibly affect the background yield.

Note that, even without complete simulations, it is understandable that, in the collinear approximation, photons conversions in the beam pipe or in the first silicon layer can be easily rejected. The opening angle of $e^+e^-$ pairs from $\gamma$ conversions at 100 MeV (worst case) is 14 mrad on average \cite{pairangle}, which, in the case of a $2$ cm distance between the conversion point and the final silicon layer, translates in a $\sim$ 280 $\mu$m distance between the hits, much larger than the required $\sim 10\,\mu$m spatial resolution.

\label{bhabha}
\subsection{Other backgrounds}
Other minor backgrounds should also be taken into account. They can easily be suppressed to small levels ($O(10^{-1})$ expected events in one month) using a proper detector design. Indeed, the most probable processes at $\sqrt{s} = 211.4$ MeV, excluding TM production and Bhabha scattering are:
\begin{itemize}
     \item $e^+e^- \to \gamma \gamma,\,e^+e^- \to \gamma \gamma \gamma$
    \item $e^+ e^- \to \mu^+\mu^-$
    \item $e^+ e^- \to \pi^0 \gamma , \,  e^+ e^- \to \pi^0 e^+ e^-$
\end{itemize}
An additional background, not linked to $e^+e^-$ interactions, is given by cosmic rays.
\subsubsection{Annihilation into gamma rays}\label{gamma}
\begin{figure}[H]
\centering
    \includegraphics[width=0.9\columnwidth]{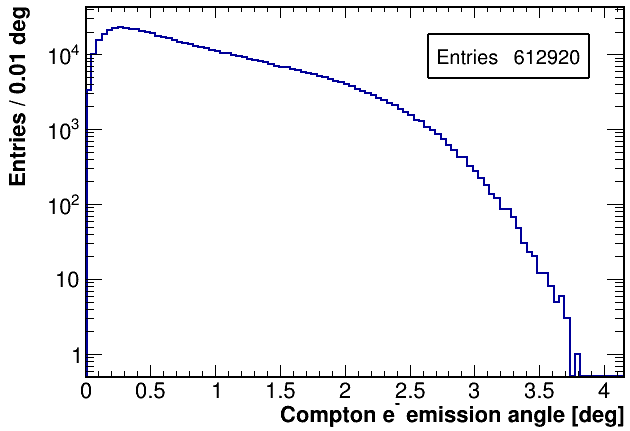}
    \caption{Distribution of the Compton electron emission angle with respect to the primary photon, for $E_\gamma = m_\mu$. Only electrons passing the cut on reconstructed energy $E > 90$ MeV, starting from $10^6$ total events, are included in the histogram.}
    \label{fig:compton}
\end{figure}

Annihilations in two or three gamma rays are very frequent at low-energy $e^+e^-$ colliders. Pair annihilation into two gamma rays (with a soft cut-off of 10 MeV on the energy of other undetected photons) has indeed a large cross-section of 5.8 $\mu$b, with the $60^\circ < \theta < 120^\circ$ angular cut already discussed \cite{photons}.\\
Photon interactions at 100 MeV can produce charged tracks that could fake electrons or positrons from signal events, when detected by both the calorimeter and all silicon layers.
If only one photon interacts in the beam pipe or in the tracker, the calorimeter cluster from the other back-to-back photon does not have any track to match, so this type events would be rejected. A signal event can instead be faked if both photons interact, through pair production or Compton scattering (with a $Z\times7.8$ mb cross section \cite{nsi}), in the 50 $\mu$m thick beryllium beam pipe 
or in the first $100$ $\mu$m thick silicon layer and only one electron/positron per photon is detected before the calorimeter. Concerning this type of events, photons undergoing conversions, accompanied by either conversion or Compton scattering of the other photon, can be suppressed at negligible levels using the same technique treated for the radiative Bhabha scattering case (see sect. \ref{bhabha}).
On the other hand, when both photons, with energy $E_\gamma \sim m_\mu$, undergo Compton scattering ($\sim 10$ events/month), suppression strategies based on energy and vertex position discrimination can be applied. If a cut at 90 MeV is applied, assuming an energy resolution of $2\%/\sqrt{E_{\mathrm{GeV}}}$ (6\% at 100 MeV), the signal efficiency is 99.4\% and this background is rejected only by a factor $\sim3.8$, using the Klein-Nishina formula. For the events passing the energy selection, the electron is emitted mainly at small angles with respect to the primary photon, as shown in Figure \ref{fig:compton}. 
A displaced vertex can be faked only if the electrons are co-planar. Then, given that 
the electrons' emission angles are over $\psi = 3.5^\circ$ with a probability of only $O(10^{-5})$ (see Figure \ref{fig:compton}), in most cases the maximum distance of any fake displaced vertex from the beam axis is $R \sin{\psi} \sim 3$ mm, where $R \sim 5$ cm is the radius of the first silicon layer.
Then, by employing a minimum cut on the displaced vertex position at 3 mm, the background from $e^+e^- \to \gamma \gamma$ is suppressed to negligible levels.

Pair annihilation into three gamma rays with energies $E_{\gamma_i}$ has a cross section of \cite{lee}:
\begin{small}
\begin{equation}
    \sigma^{3\gamma}(E_{\gamma_i} > k \frac{\sqrt{s}}{2}) = \frac{2 \alpha^3}{s} \left [3 - \frac{2\pi^2}{3} - (\ln{4 \gamma^2} - 1)^2 (2 \ln{k} + 1) \right]
\end{equation}
\end{small}
where $k$ is the relative soft cut-off for all three photons and $\gamma = \frac{\sqrt{s}}{2m_e}$. With $k = 0.1$ ($\sim$ 10 MeV cut-off), the resulting cross section is 3 $\mu$b. When two photons are collinear and the calorimeter sees two clusters with reconstructed energy greater than 90 MeV, as in the two-photons case, 
the only non-negligible background is due to double Compton scattering.
Indeed, when at least one $\gamma$ conversion occurs, the event is rejected as for radiative Bhabha scattering, while the remaining case, i.e. triple Compton scattering, not only has a negligible yield but can also be discarded easily by distinguishing the three electron tracks.

In the case of double Compton scattering, the isolated photon has indeed reconstructed energy over 90 MeV, therefore the emitted electron angle is mostly contained in the $3.5^\circ$ range as before, while the other photon undergoing Compton scattering can also have smaller energy and produce electrons at wider angles. 
An angular cut on the opening angle $\alpha > 177^\circ$ between the two tracks, as reconstructed by the silicon detector, is applied (see sect. \ref{bhabha}), therefore the sum of electrons emission angle is bounded under $3^\circ$, in the hypothesis that one electron is emitted clockwise and the other counterclockwise. 
A displaced vertex cannot be geometrically faked, indeed, if both electrons are emitted clockwise, or vice versa.
After the cut $\alpha > 177^o$, the annihilation in three photons can be treated as the two photons case, therefore its background yield is negligible.\\
A more quantitative study of the two and three-photon background, particularly from the point of view of the tracker response, requires full detector simulations which are outside the scope of this article.

\subsubsection{Muon pair production}
\begin{figure}[H]
\centering
    \includegraphics[width=0.9\columnwidth]{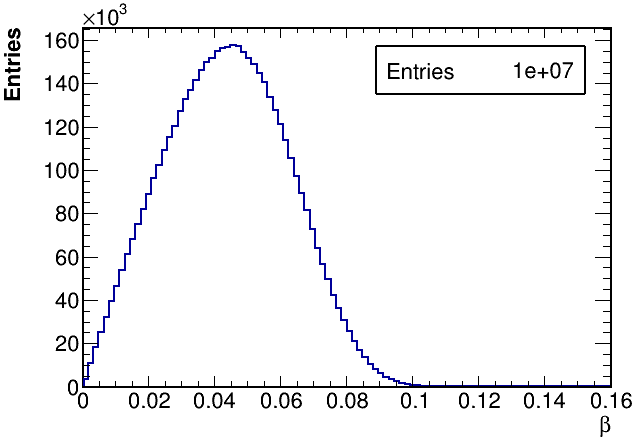}
    \caption{Muons $\beta$ from simulation.}
    \label{fig:betamuons}
\end{figure}

Muon pair production takes place about the threshold, with cross section given by \cite{Brodsky:2009gx}
\begin{equation}
    \sigma(e^+e^-\to\mu^+\mu^-) = \frac{2\pi \alpha^2 \beta (3 - \beta^2)} {3s} S(\beta)\,,
\end{equation}
where the Sommerfeld-Sakharov-Schwinger \cite{sommerfeld, Schwinger:1973rv, Sakharov} enhancement factor
\begin{equation}
    S(\beta) = \frac{X(\beta)}{1-e^{-X(\beta)}} \quad \text{with} \quad X(\beta) = \frac{\pi \alpha}{\beta}\sqrt{1-\beta^2}
\end{equation}
has been included. In the above, $\beta = \sqrt{1- \frac{4m_\mu^2}{s}}$ is the speed (divided by $c$) of the outgoing muons \cite{mp2}. The muons are produced only when $\sqrt{s} > 2m_\mu$, so it must be taken into account the beam energy spread ($\sigma_E/E \sim 2\times 10^{-3})$ \cite{bes}. The cross-section can then be evaluated by simulation, extracting random values of $\sqrt{s}$ from a Gaussian distribution centered on 2$m_\mu$ with a $\sigma_{\sqrt{s}} = 300$ keV, and calculating the resulting values of $\beta$, as seen in Figure~\ref{fig:betamuons}.\\
The average cross-section (without angular cuts) is evaluated to be 164 nb, which is about 3000 times higher than the 62 pb cross-section for TM production but about 100 times less than Bhabha scattering. 
The produced muons have a kinetic energy of less than a few MeV, so they decay inside the beam pipe material or the tracker, and their energy deposit in the calorimeter is due to the Michel electron from the decay, or to photons from radiative muon captures (RMC), for $\mu^-$ only. 

Positive muons can only decay freely, dominantly via the Michel \cite{pdgmichel} mode $\mu \rightarrow e+ \nu_{\mu}+ \bar{\nu_{e}}$, having a kinematic endpoint for the outgoing electron of $E_{max} = (m_{\mu}^2+m_{e}^2)/(2m_{\mu}) = 52.8$ ~MeV, with a spectrum parametrized as follows:
\begin{equation}
    \frac{d \Gamma}{dx} \propto 3x^2 - 2 x^3
\end{equation}
where $x=E/E_{max}$.

These electrons can be suppressed at negligible levels with the same calorimetric energy cut at 90 MeV employed for the two-photons case (see sect. \ref{gamma}).
\subsubsection{Neutral pion production}
The reaction $e^+e^- \to \pi^0 \gamma$ has a cross-section of \cite{pion}:
\begin{equation}
 \sigma(e^+e^- \to \pi^0 \gamma) = \frac{8 \,\alpha\, \Gamma_{\pi^0\to\gamma\gamma}}{3 m_{\pi^0}^3} \left (  1  - \frac{m_{\pi^0}^2}{s} \right )^3
\end{equation}
which corresponds to a value of 5 pb. There are three emitted photons (one as a recoil photon and two from $\pi^0$ decay) so this background is suppressed to negligible levels as in the three-photons case.\\
Another process producing neutral pions is $e^+e^- \to \pi^0 e^+e^-$, via photon-photon fusion, with a cross-section of 22 pb \cite{pion}.
If the calorimeter detects four (two photons from $\pi^0$ decay and an electron-positron pair) or three different clusters, the events are directly discarded. If only two clusters are detected, because the photons from $\pi^0$ are emitted back-to-back and are collinear with electron or positron tracks, two situations must be distinguished. If there are photon interactions before the calorimeter, the technique treated for radiative Bhabha scattering (see sect. \ref{bhabha}) can be applied to partially suppress this type of events. Otherwise, in the case of uninteracting or undetected photons, it is possible to see only two tracks in the silicon detector associated with two clusters in the calorimeter with energy compatible with $\sqrt{s}/2$, so this background is simply treated as very small contamination to Bhabha scattering, with a relative yield less than the ratio of the $e^+e^- \to \pi^0 e^+e^-$ to the Bhabha scattering cross section: $\sim$ 22 pb / 9 $\mu$b $ = 2.5 \times 10^{-6}$.

\subsubsection{Cosmic rays}
Cosmic ray events can fake a signal event when they are not rejected by the calorimeter using cluster shapes or energy information. In the region inside the first silicon layer (with a radius of 5 cm), the number of cosmic ray events expected in one month is $O(10^7)$, assuming a 20 cm cylinder length. \\
This background can be rejected using a longitudinally segmented and high granularity crystals calorimeter. A possible solution is a cylindrical barrel calorimeter around the beam pipe as the one proposed for the future Muon Collider, Crilin \cite{CrilinCalo},\cite{Daniele}, with LYSO crystals and SiPM photo-sensors readout. Cluster shape and deposited energy analysis allows clear discrimination between cosmic rays and back-to-back $e^+e^-$ pairs coming from beam interactions. 
This kind of calorimeter design allows discrimination between electron clusters and cosmic ray tracks, based on cluster shape, deposited energy topology, and time-of-flight characteristics. Hence, high muon rejection can be achieved by exploiting the high granularity, segmentation, and excellent timing capabilities.
In order to increase the calorimeter discrimination factor by about 4 orders of magnitude, a hermetic cosmic ray veto detector surrounding the calorimeter can be built, leading to an expected number of cosmic rays events to be kept within $10^{-1}$ in one month.

\section{Energy scan and off-peak measurements}
By performing dedicated measurements in off-signal regions, for $\sqrt{s}$ above and below the TM mass, a data-driven characterization and subtraction of the expected background can be carried out, with a focus on the evaluation of the machine background. Indeed, although the machine parameters were extrapolated from DA$\Phi$NE performances at nominal energy, a complete simulation of the interaction zone is lacking, as it is the determination of the beam background. As a result, the aforementioned data-driven solution can conservatively provide a good indication of the background contamination at the TM $\sqrt{s}$, assuming that the continuum background distribution near the TM peak is flat.

Furthermore, dedicated energy scans around the $\mu^+ \mu^-$ production threshold ($2m_\mu = 211.4$ MeV), can provide an absolute indication of the center-of-mass (CM) energy, against the sharp increase in $\mu^+ \mu^-$ production, which can be evaluated based on the reconstruction of the previously described Michel spectra for electrons from muon decays.

\section{Discovery potential}
\label{dp}
The number of background events expected in one month of data-taking has two main contributes: 1) the Bhabha scattering which can be suppressed using analysis cuts, and 2) the cosmic rays events that are independent on $D_r$ and $x_v$, corresponding to less than 0.64 events/month.\\
The number of signal events has been estimated with the same detector simulation used for Bhabha scattering, and the secondary vertex has been simulated with contributions from all excited states.
In order to achieve an effective signal/background discrimination, the following selection is required:
\begin{itemize}
    \item Pre-selection cuts:
        \begin{itemize}
            \item two reconstructed tracks with associated clusters in the calorimeter;
            \item two calorimeter clusters with an energy greater than 90 MeV (see sect. \ref{gamma});
            \item two and only two opposite charged particles detected in all silicon layers (see sect. \ref{bhabha});
            \item opening angle greater than 177$^\circ$ (see sect. \ref{gamma});
            \item vertex $x_v$ coordinate less than 40 mm
            \item $D_r < $ 44 mm
        \end{itemize}
    \item Analysis cuts:
        \begin{itemize}
            \item Circular cut (see eq. \ref{eq:circ}): required events out of the circle with center (0 mm, 4 mm) with 6 mm radius in the plane $(x_v, \, D_r)$
            \item $D_r > $ 5 mm
        \end{itemize}
\end{itemize}

\begin{figure}[H]
\centering
    \includegraphics[width=\columnwidth]{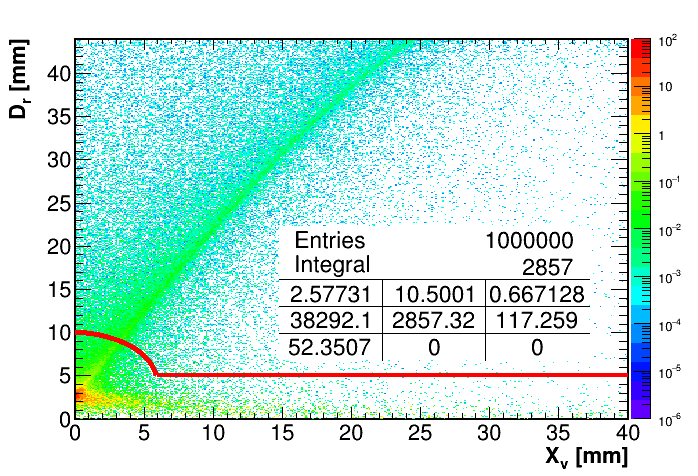}
    \caption{2D distribution of $x_v$ and $D_r$ for signal events. The red lines represent the analysis cuts (see Figure \ref{fig:bhabhath2}).}
    \label{fig:signal}
\end{figure}
\begin{figure}[H]
\centering
    \includegraphics[width=\columnwidth]{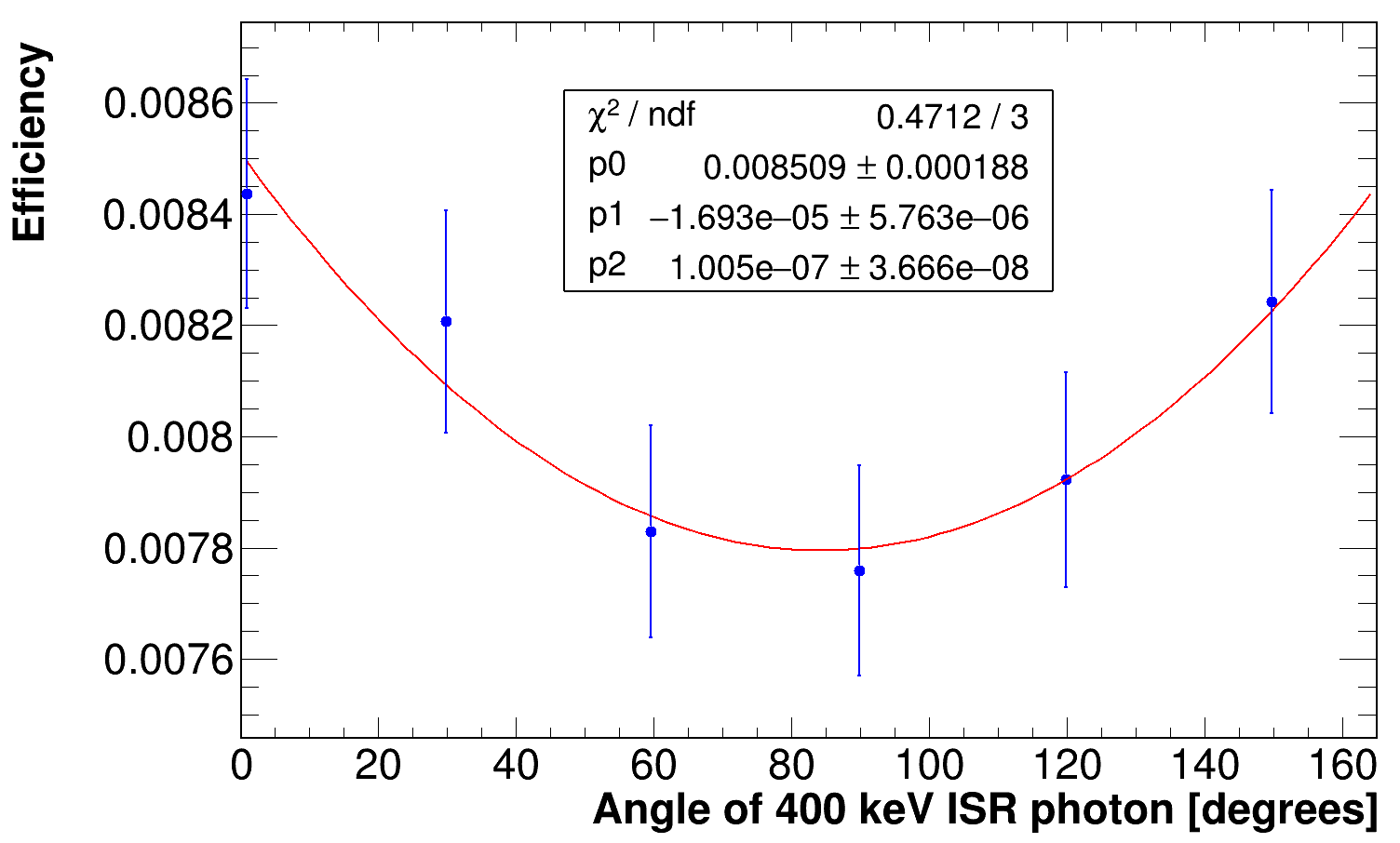}
    \caption{Efficiency of the signal including a 400 keV ISR photon for all the events with different polar angles. The values have been fitted with a parabola to extract the minimum value.}
    \label{fig:isreff}
\end{figure}

The joint distribution of $D_r$ and $x_v$ distribution for signal events, not including ISR effects, after pre-selection cuts, is shown in Figure \ref{fig:signal}.
The upper cuts on vertex $x_v$ coordinate (40 mm) and $D_r$ (44 mm) are due to the presence of the beam pipe at a radius of 44 mm.
The efficiency on the signal after pre-selection and analysis cuts is $(8.2 \pm 0.2) \times 10^{-3}$, without including ISR effects.
As shown in Figure \ref{fig:isr}, the 99.5\% of the events includes an ISR emission with an energy lower than 400 keV. In order to roughly estimate the effect of ISR on the signal, the presence of an ISR photon with a 400 keV energy has been included for different values of the ISR photon polar angle and with uniformly distributed azimuthal angles, as shown in Figure \ref{fig:isreff}.
The efficiency of the signal can then be underestimated as $7.8 \times 10^{-3}$, looking at the lowest efficiency in Figure \ref{fig:isreff}. The expected number of signal events in 30 days of data-taking with a 10 pb$^{-1}$/day luminosity and a 39 pb cross section is then greater than 91. Given that the expected number of background events is less than 0.64, the significance is greater than 27$\sigma$, greatly exceeding the conventional 5$\sigma$ threshold for discovery.
It has been also established that, with the same cuts and the same data-taking time, a discovery can be achieved with luminosity values as low as 0.5 pb$^{-1}$/day, as the expected number of signal events, background from cosmic rays and Bhabha scattering are, respectively, 4.6, 0.1 (unvaried) and 0.03.
\section{True Muonium spectroscopy}

As already mentioned, one of the motivations for the study of TM is its potential in probing BSM physics or precision SM physics, and especially in shedding new light on the longstanding muon $g-2$ ($ a_\mu$) problem. Indeed, precise measurements of the properties of TM (hyper-fine splittings, Lamb shift) can be used in combination with measurements of $a_\mu$ to probe several BSM scenarios in which the new physics couples to muons \cite{TM-LHCb}, as they in general predict modifications 
of these properties, and also to estimate the contribution to $(g-2)_\mu$ from hadronic vacuum polarization \cite{PhysRevA.95.012505}.
As explained in the introduction, TM has a broad physics reach, due to the absence of large hadronic contributions and to the large reduced mass compared to positronium or muonium.


In an experimental phase subsequent to discovery, if a large enough number of TM bound states is available, the Lamb shift may be measured by means of a laser of appropriate frequency to excite the $P$ states of TM. Focusing on the $n=2$ states, a laser frequency of about $10$ THz \cite{refId0} would excite $2 S$ states to $2 P$, which, as discussed above, would then have to decay to $1S$ emitting X-rays in the keV range, so measuring the $2P \to 1S$ transitions X-ray yield as a function of the laser frequency provides an estimate of the Lamb shift.
BSM contributions modify the Lamb shift frequency by $O(100)$ MHz, a value in the reach of modern spectroscopy techniques, therefore TM could be efficiently used as a BSM probe.
Finally, it should be remembered that the measurement of the hydrogen atom Lamb shift was a milestone of modern physics, as it confirmed QED correctness, and that bound states spectroscopy, in general, provides very sensible probes. 
\section{Conclusion}
Several of BSM discovery potential is hidden in the atomic spectroscopy of QED-bound states. Among these interesting objects, one of the most sensible ones, is the so-called true muonium (TM), a $\mu^+\mu^-$ bound state. TM can be produced on resonance with a 67 nb cross-section at $e^+e^-$ collider running at a $2m_\mu = 211.4$ MeV center-of-mass energy.\\
In real-world scenarios, due to the smallness of its B.E. with respect to the beam's energy spread $\sigma_E$, the cross-section is smaller by a factor proportional to $\sigma_E$.\\
In this paper, the DA$\Phi$NE collider at the Frascati National Laboratory of INFN \cite{lnf} is used, which now runs at $\sqrt{s} = 1020$ MeV, as the benchmark of machine requirements for TM research. For this reason, its beam and collision parameters were studied to assess whether there is a potential for discovery of TM excited states, in the hypothesis to run at the proper $\sqrt{s}$. The TM decay vertex to an electron pair can indeed be reconstructed and employed to discriminate signal over background. It was therefore shown that TM excited states can be observed in a data-taking of the order of one month using a detector with a polar acceptance angle of $\theta=60^\circ$, made of a multi-layer silicon tracker with 10 $\mu$m spatial resolution, a calorimeter with a resolution better than $2\%/\sqrt{E[\mathrm{GeV}]}$, and a hermetic cosmic ray veto.\\
Previous proposals for TM discovery involved the construction of $e^+e^-$ colliders with large collision crossing angles ($\theta \sim 30^\circ)$, in order to provide TM with enough boost to observe its fundamental state \cite{dimus}. On the contrary, it was proved that also the small crossing angle of already existing colliders like DA$\Phi$NE is sufficient to discover TM, by observing its excited states.

\section{Acknowledgements}
The authors are grateful to M. Raggi and S. Miscetti, for the careful reading of the paper and for valuable suggestions and comments. They also thank the whole Acceleration Division of the Laboratori Nazionali of Frascati for providing them with the DA$\Phi$NE beam parameters used in this paper and C. Carloni Calame for help with the Babayaga generator.

\appendix
    \section{Methods}
\label{methods}
The significance is calculated as: 
\begin{equation}
    Z = \sqrt{-2 \log{L(N, 0)/L(N, 1)}}
\end{equation}
where $L(N, \mu)$ is the Poissonian likelihood with $N$ observed events, a signal strength of $\mu$ (0 for background only, 1 for nominal signal yield), $s$($b$) expected signal (background) events \cite{pratstat}
\begin{equation}
    L(n, \mu) = \frac{(\mu s + b)^N}{N!} \exp{-(\mu s + b)}
\end{equation}

\section{Initial State Radiation} \label{app:isr}
The ISR radiator function used in Eq. \eqref{eq:isr-rate} and Eq. \eqref{eq:isr-boost-weight} is essentially the probability that the electron pair carries a given fraction of the nominal center of mass energy. The following relations were used: \cite{greco, jadach1, jadach2}
\begin{equation}
f_{\text{ISR}}(x;s) = f^0_{\text{ISR}}(x; s)\,\left(1 + \frac{\beta_l}{2} - \frac{1}{2}(1-x^2)\right)\,,
\end{equation}
where $\beta_l = \frac{2\alpha}{\pi}\left(\log{\frac{s}{m_e}} - 1 \right)$, and
\begin{equation}
f^0_{\text{ISR}}(x;s) = \frac{\exp{\left(\frac{\beta_l}{4} + \frac{\alpha}{\pi}\left(\frac{1}{2} + \frac{\pi^2}{3}\right)-\gamma_E \beta_l\right)}}{\Gamma(1+\beta_l)} \beta_l(1-x)^{\beta_l - 1}\,.
\end{equation}

\bibliography{apssamp}

\end{document}